\documentclass[useAMS,usenatbib,usegraphicx]{mn2e}
\usepackage{amssymb}

\def\fits{\textsc{fits}}
\def\iraf{\textsc{iraf}}
\def\ccdproc{\textsc{ccdproc}}
\def\daophot{\textsc{daophot}}
\def\echelle{\textsc{echelle}}
\def\makee{\textsc{makee}}
\def\redux{\textsc{hires redux}}
\def\molly{\textsc{molly}}
\def\doppler{\textsc{doppler}}
\def\elc{\textsc{elc}}
\def\ut{\textsc{ut}}

\newcommand\ion[2]{#1$\;${\scshape{#2}}}

\title[Optical observations of HETE~J1900.1$-$2455]{Optical photometry and spectroscopy
       of the accretion-powered millisecond pulsar HETE~J1900.1$-$2455}
\author[P. Elebert et al.]{P. Elebert,$^{1}$\thanks{E-mail: p.elebert@ucc.ie}
        P. J. Callanan,$^{1}$
        A. V. Filippenko,$^{2}$
        P. M. Garnavich,$^{3}$
        G. Mackie,$^{4}$ 
        \newauthor J. M. Hill$^{5}$ and
        V. Burwitz$^{6}$\\
$^{1}$Department of Physics, University College Cork, Cork, Ireland\\
$^{2}$Astronomy Department, University of California, Berkeley, CA 94720--3411, USA\\
$^{3}$Department of Physics, University of Notre Dame, Notre Dame, IN 46556--5670, USA\\
$^{4}$Centre for Astrophysics and Supercomputing, Swinburne University of Technology, Hawthorn, Victoria 3122, Australia\\
$^{5}$Large Binocular Telescope Observatory, University of Arizona, Tucson, AZ 85721--0065, USA\\
$^{6}$Max-Planck-Institut f\"{u}r extraterrestrische Physik, Giessenbachstra{\ss}e, D-85741 Garching, Germany}

\begin{document}


\pagerange{\pageref{firstpage}--\pageref{lastpage}} \pubyear{2008}

\maketitle

\label{firstpage}


\begin{abstract}
   \noindent
   We present phase resolved optical photometry and spectroscopy of the accreting millisecond pulsar
   HETE~J1900.1$-$2455. Our $R$-band light curves exhibit a sinusoidal modulation, at close to the
   orbital period, which we initially attributed to X-ray heating of the irradiated face of the
   secondary star. However, further analysis reveals that the source of the modulation is more
   likely due to superhumps caused by a precessing accretion disc. Doppler tomography of a broad
   H$\alpha$ emission line reveals an emission ring, consistent with that expected from an accretion
   disc. Using the velocity of the emission ring as an estimate for the projected outer disc
   velocity, we constrain the maximum projected velocity of the secondary to be 200 km~s$^{-1}$,
   placing a lower limit of 0.05 M$_{\sun}$ on the secondary mass. For a 1.4 M$_{\sun}$ primary,
   this implies that the orbital inclination is low, $\lesssim$ 20\degr. Utilizing the observed relationship
   between the secondary mass and orbital period in short period cataclysmic variables, we estimate
   the secondary mass to be $\sim$0.085 M$_{\sun}$, which implies an upper limit of $\sim$2.4
   M$_{\sun}$ for the primary mass.
\end{abstract}


\begin{keywords}
   accretion, accretion discs --
   techniques: photometric --
   techniques: spectroscopic --
   pulsars: individual: HETE~J1900.1$-$2455 --
   X-rays: binaries
\end{keywords}


\section{Introduction}

Low mass X-ray binaries (LMXBs) are close binary systems consisting of a degenerate primary -- black
hole or neutron star (NS) -- accreting matter from a low mass ($<$ 1 M$_{\sun}$) secondary, via Roche
lobe overflow. X-ray transients (XRT) are LMXBs which undergo periodic outbursts, explained by the disc
instability model \citep[e.g.][]{dubus2001}, while persistently bright systems are permanently in the outbursting
state. One subclass of LMXBs are the accretion-powered millisecond pulsars
(AMSPs), the first of which, SAX J1808.4$-$3658, was discovered in 1998 \citep{wijnands1998}. Seven more of
these systems have been discovered since then \citep{poutanen2006, wijnands2006, markwardt2007}.

\vspace{1mm}

In general, the outburst duration for AMSPs ranges from a few weeks to months.
AMSPs generally
contain a weakly magnetized ($\sim$10$^8$--10$^9$ G) NS with spin frequencies between 180 and 600 Hz. The
orbital periods range between 40 min and 5 h. The secondary star in these systems is either a white or brown dwarf
\citep[][and references therein]{falanga2007}. \cite{falanga2005} have measured the predicted decrease in the NS
spin period for the AMSP system IGR J00291$+$5934, supporting the idea that these AMSPs are in fact old NS, which
have over time been spun up to millisecond periods by acquiring angular momentum from the accretion of material
from the secondary.
As such, they provide the missing link between LMXBs and old, isolated millisecond radio pulsars. 

\vspace{1mm}

In comparison to brighter LMXBs, relatively little is known about the optical properties of AMSPs. Even in outburst,
the eight so far discovered are comparatively faint, the brightest being SAX J1808.4$-$3658 at $R$ $\simeq$ 16.2 mag
\citep{wang2001}. In quiescence, these systems are generally very faint optically, the
brightest measured again being SAX J1808.4$-$3658 at $R$ $\simeq$ 20.9 mag.

\vspace{1mm}

HETE~J1900.1$-$2455 is the seventh AMSP, discovered by \cite{vanderspek2005} on 2005 June 14 by the
{\textit{High Energy Transient Explorer II}} ({\textit{HETE II}}).
Several type-I bursts have since been observed, and assuming that
the peak flux observed during the brightest of these is
Eddington-limited, HETE~J1900.1$-$2455 is at a distance of $\sim$4.3 kpc \citep{suzuki2007}.

\vspace{1mm}

The pulsation frequency was determined with the {\textit{Rossi X-ray Transient Explorer}} to be 377.3 Hz \citep{morgan2005}.
Pulse timing analysis by \cite{kaaret2006} revealed a circular orbit with a period ($P_{\mathrm{orb}}$) of $\sim$83.3
minutes (4995.258 $\pm$ 0.005 s), and a projected primary semimajor axis, $a_{1}\mathrm{sin}i$,
of 18.41 $\pm$ 0.01 $\times$ 10$^{-3}$ light second. These parameters combine to provide a mass function
$f(M) = 2.004 \pm 0.003 \times10^{-6}\ \mathrm{M_{\sun}}$.
\cite{galloway2006} report that pulses were detected intermittently in the first two months after
discovery. No pulses have been reported since then.

\vspace{1mm}

Unusually for an XRT, this system has remained active for more than 2 years, at a level of $\sim$2 $\times$ 10$^{36}$
erg s$^{-1}$. Several times during
2007 March/February, the X-ray flux fell by an order of magnitude, but each time
returned to outburst levels within $\sim$1 week \citep{galloway2007}. In 2007 May, the source was observed to decline to
its lowest level since discovery (by a factor of $>$ 10$^{3}$), lasting for several weeks, prompting speculation that
HETE~J1900.1$-$2455 was returning to quiescence \citep{degenaar2007a,galloway2007,torres2007}. However, the system
returned to its outburst state within $\sim$2--3 weeks \citep{degenaar2007b,garnavich2007}.

\vspace{1mm}

The optical counterpart was found by \cite{fox2005} with an $R$-band magnitude of $\sim$18.4 mag.
Subsequent observations by \cite{steeghs2005} found
the $R$-band magnitude to be 18.02 $\pm$ 0.03 mag, and the $V-R$ colour (dereddened) to be $-0.16$ mag, with
spectroscopy revealing a broad \ion{He}{ii} $\lambda$4686 \AA\ emission line.

\vspace{1mm}

In this paper we present optical photometry and spectroscopy of HETE~J1900.1$-$2455 in its outburst state. These
observations were motivated by several goals, key amongst them the aim of measuring the radial velocity of the
secondary, and hence constraining the mass of the compact object. For the longer period transient systems, such
measurements are possible via radial velocity studies of absorption lines from the secondary once the system has
returned to quiescence. For the shorter period AMSPs however, the secondary is likely to be so faint as to make
such observations extremely difficult, if not impossible. However, \cite{steeghs2002} and \cite{casares2004} have
shown how radial velocity measurements of the secondary are still possible for X-ray bright systems, by studying the
Bowen blend emission from the irradiated face of the secondary: this technique may offer the only way of measuring
the velocity of the secondary in AMSPs (i.e. whilst they are still X-ray bright), and usefully constraining the
mass of the neutron star.


\section{Data}

Our data consist of photometry obtained with the 2.3~m Advanced Technology Telescope (ATT) at Siding Spring
Observatory, Australia, the 3.5~m Wisconsin-Indiana-Yale-NOAO (WIYN) telescope at Kitt Peak, Arizona, and the
8.4~m Large Binocular Telescope (LBT) at Mount Graham, Arizona. Our spectroscopy was obtained using the HIgh
Resolution \'{E}chelle Spectrometer (HIRES) on the 10~m Keck~I telescope at Mauna Kea, Hawaii.

\subsection{Photometry}

We obtained photometry on 2 nights in 2006 September with the 2.3~m ATT, using the CCD4240 e2v detector, mounted at
the $f/18$ Nasmyth focus. The detector has 2148 $\times$ 2148 13.5 $\umu$m pixels, with 0.\arcsec34~pixel$^{-1}$.
On 2006 September 21 \ut, we obtained 16 $R$-band exposures, covering slightly more than 1 orbital period, with
exposure times of 60 s ($\times\ 1$), 300 s ($\times\ 9$) and 240 s ($\times\ 6$). On the following night,
we obtained $R$, $V$ and $B$-band exposures, 2 $\times$ 300 s exposures of each. On both nights, exposures were
taken of the photometric standard field SA-112 \citep{landolt1992}. Seeing was $\sim$2\arcsec\ on both nights, but
deteriorated towards the end of the second night.

Three 300 s $R$-band exposures were also taken with the 3.5~m WIYN telescope on 2006 September 19 \ut, using the
WIYN Mini-Mosaic Imager, consisting of 2 SITe CCDs, each with 4096 $\times$ 2048 15 $\umu$m pixels, with
0.\arcsec14~pixel$^{-1}$. The seeing was $\sim$2\arcsec\ for all 3 exposures. Exposures were also taken of the
standard star field PG031$+$051 \citep{landolt1992}.

\begin{figure}
  \begin{center}
    \includegraphics[scale=0.44, angle=0]{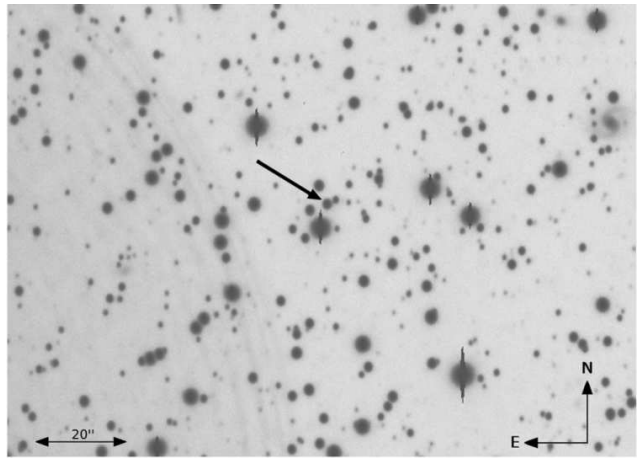}
  \caption{LBT image of the field of HETE~J1900.1$-$2455 (Sloan $r$-band), 2.\arcmin3 $\times$ 1.\arcmin6. The black
           arrow indicates the position of HETE~J1900.1$-$2455.}
  \label{1900}
  \end{center}
\end{figure}

After the decline/rebrightening episode of 2007 May/June, we obtained additional observations using the LBT.
On 2007 June 13 \ut\ and June 24 \ut\ we obtained
200 s Sloan $r$-band exposures of HETE~J1900.1$-$2455 with the Large Binocular Camera (Blue) 
(Giallongo et al., in preparation; \citealt{ragazzoni2006}) at the prime focus of the
LBT (20 exposures per night). The eev-blue detector consists of 4 CCDs with 2048 $\times$ 4608 13.5 $\umu$m
pixels, with 0.$\arcsec$224~pixel$^{-1}$. The seeing on 2007 June 13 was $\sim$0.\arcsec85--1.\arcsec15, while on
June 24 was $\sim$1\arcsec--1.\arcsec5.
The HETE~J1900.1$-$2455 field is shown in Fig. \ref{1900} (from LBT on 2007 June 13, with seeing $\sim$0.\arcsec85).

All the photometry data were reduced in the same way. The frames were firstly bias corrected, trimmed and
flat-fielded using the \ccdproc\ routines in \iraf\footnote{\iraf\ is distributed by the National Optical
Astronomy Observatories, which are operated by the Association of Universities for Research in Astronomy, Inc.,
under cooperative agreement with the National Science Foundation.}.
Photometry was performed using the
\daophot\ \citep{stetson1992} point spread function fitting package in \iraf.
The magnitude of HETE~J1900.1$-$2455 was measured relative to several bright stars.
These relative magnitudes were then calibrated by comparison with the standard star frames.
Note that the WIYN data was used for photometric calibration only.

The ATT data from 2006 September gives $R = 17.91 \pm 0.05$ mag, $V = 18.00 \pm 0.04$ mag and $B = 17.92 \pm 0.04$ mag.
The errors quoted are an estimate of the systematic uncertainty.
We note that the measured $R$-band magnitude is
consistent with that reported by \cite{steeghs2005}. On 2007 June 13, $R = 18.51 \pm 0.05$ mag, while on 2007 June 24,
$R = 18.44 \pm 0.05$ mag.

Taking the value for the hydrogen column density to the source, $N_{\mathrm{H}}$, as
1.6 $\pm$ 0.4 $\times$ $10^{21}$ cm$^{-2}$ \citep{campana2005}, and using the method detailed by
\cite{savage1979} yields $E(B-V) = 0.30 \pm 0.08$. 
The resulting dereddened colours are ($B-V$)$_0 = -0.34$ $\pm$~0.08 mag, and ($V-R$)$_0 = -0.17$ $\pm$~0.06 mag
\citep[again, consistent with the value reported by][]{steeghs2005}.

In the $R$-band, HETE~J1900.1$-$2455 is blended with a faint star 2\arcsec\ to the northwest (see Fig. \ref{1900}).
This star was fitted simultaneously with HETE~J1900.1$-$2455, and was found to have an $R$-band magnitude of 20.1
$\pm$ 0.1 mag, which was constant, within these errors, in all our observations. This star was not visible in the $B$
or $V$-band frames. This is consistent with the observations of \cite{steeghs2005}, who note that this blended star
is $\sim$2 magnitudes fainter (in the $R$-band) than HETE~J1900.1$-$2455.

In the LBT data, in frames with good seeing, an additional faint star is observed $\sim$1\arcsec\ northwest of HETE
J1900.1$-$2455, with an $R$-band magnitude $\sim$5 mag fainter than HETE~J1900.1$-$2455.

\begin{figure}
  \begin{center}
    \includegraphics[scale=0.245, angle=-90]{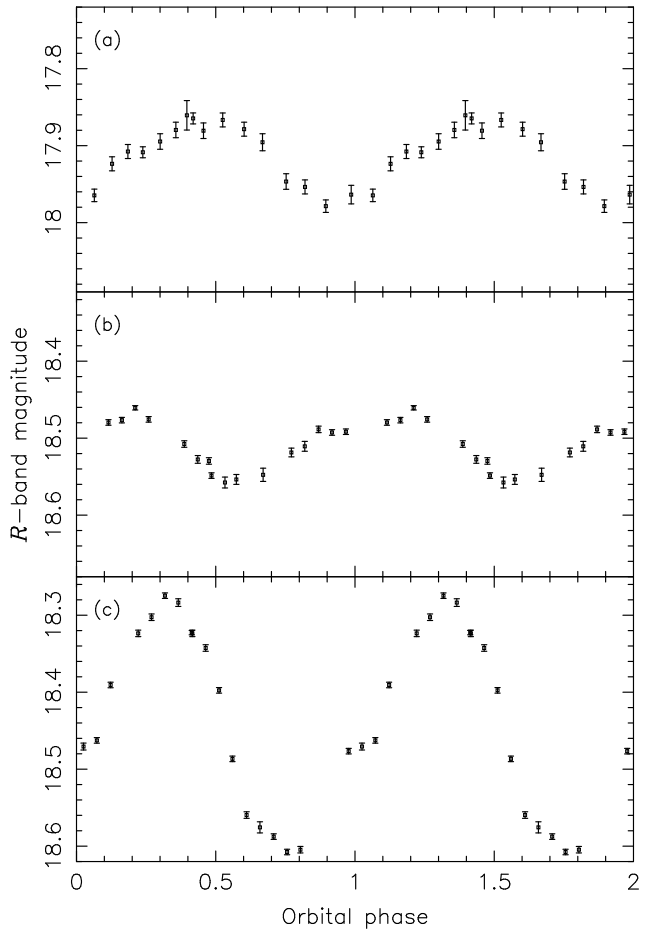}
  \caption{$R$-band light curves, folded on the pulse timing ephemeris \citep{kaaret2006}, and plotted twice for
           clarity. (a) ATT light curve from 2006 September 21, (b) LBT light curve from 2007 June 13, (c) LBT light
           curve from 2007 June 24.}
  \label{lc}
  \end{center}
\end{figure}

In Fig. \ref{lc} we plot the three $R$-band light curves, folded on the pulse timing ephemeris of \cite{kaaret2006}.
The phase error at the epochs of these optical observations, extrapolated from the the error in the period of the pulse
timing ephemeris, is $\sim$0.01. Each light curve exhibits a similar modulation, but that of 2007 June 24 has an amplitude
$\sim$3 times greater than for the other two. Clearly, the phasing of the three light curves is not consistent,
implying that the period of this modulation is different to the orbital period.

\subsection{Spectroscopy}

Our spectroscopy was obtained using the HIRES spectrometer \citep{vogt1994} at the right Nasmyth focus of the
10 m Keck~I telescope, with a 52.68 lines mm$^{-1}$ \'{e}chelle grism and the HIRESr 250 lines mm$^{-1}$ long
wavelength cross disperser. The slit width was 0.\arcsec86 with a slit length of 7\arcsec. The \'{e}chelle grating
angle was 0$\degr$ and the cross disperser grating angle was 0.\degr2405, providing a wavelength coverage of
$\sim\lambda\lambda$3900--8500 \AA, spread over a mosaic of 3 MIT-Lincoln Labs CCDs (``red'', ``green'' and ``blue''),
each with 2048 $\times$ 4096 15 $\umu$m pixels, with 0.\arcsec12~pixel$^{-1}$ in the spatial dimension. The images
were binned by 2 pixels in the spatial dimension.

15 \'{e}chelle spectra were obtained with HIRES on 2006 October 11 \ut. Appropriate bias, flat, trace star,
ThAr arclamp and flux standard (G191B2B) exposures were also taken. The object exposure times were 300 s
($\times\ 5$), 360 s ($\times\ 8$) and 420 s ($\times\ 2$), with the 15 spectra covering slightly more than
one orbital period.

The spectra were extracted using the \makee\footnote{http://spider.ipac.caltech.edu/staff/tab/makee/}
(MAuna Kea \'{E}chelle Extraction) HIRES reduction package, which takes the two dimensional
raw \fits\ images, and produces optimally extracted, wavelength calibrated, one dimensional spectra.
The lines of interest to us were on the ``blue'' and ``green'' CCDs.
The wavelength calibration was performed using the ThAr arc lamp exposures, by fitting a sixth order
polynomial to an average of 30 lines per \'{e}chelle order (minimum 20); this resulted in a dispersion of
0.015--0.025 \AA~pixel$^{-1}$ for the ``blue'' CCD and 0.021--0.032 \AA~pixel$^{-1}$ for the
``green'' CCD, and an RMS scatter of $<$ 0.003 \AA\ (average $\sim$0.002 \AA).
To confirm the extraction method implemented in \makee,
the spectra were also extracted using the \redux\ software\footnote{http://www.ucolick.org/$\sim$xavier/HIRedux/},
and the \echelle\ package in \iraf: however the spectra extracted using all three methods were very similar.

\begin{figure}
  \begin{center}
    \includegraphics[scale=0.36, angle=270]{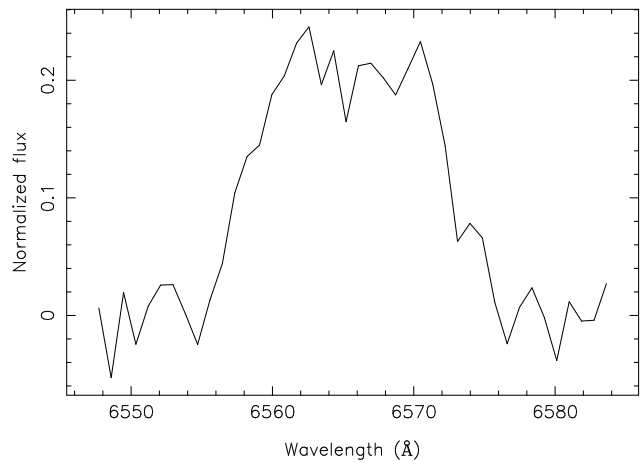}
  \caption{Average spectrum for $H\alpha$, normalized to the local continuum.}
  \label{ha_avg}
  \end{center}
\end{figure}

\begin{figure}
  \begin{center}
    \includegraphics[scale=0.36, angle=270]{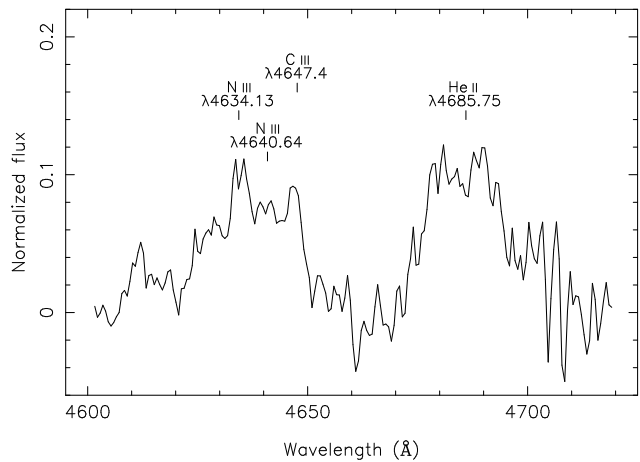}
  \caption{Average spectrum for the Bowen blend and \ion{He}{ii} $\lambda$4686~\AA, normalized to the local continuum.
           The spectrum has been shifted to the rest frame of the system, based on the systemic velocity determined
           from the H$\alpha$ emission line in section 3.2. The spectrum has also been boxcar smoothed.}
  \label{bowen_heII_avg}
  \end{center}
\end{figure}

The main emission features present in our spectra are H$\alpha$ $\lambda$6563 \AA, \ion{He}{ii}
$\lambda$4686 \AA\ and the Bowen blend of \ion{N}{iii} and \ion{C}{iii} lines centred at $\sim$$\lambda$4640 \AA.
H$\beta$ is present as a weak emission line within a broader absorption feature.
The \'{e}chelle orders containing these features were exported
for further processing with the \molly\ spectroscopic data analysis package, and \doppler,
a package used for performing maximum entropy Doppler tomography.
The spectra were re-binned onto a velocity scale of 40 km s$^{-1}$~pixel$^{-1}$,
yielding a signal-to-noise (S/N) ratio of 6--16 pixel$^{-1}$. The S/N varied across each order, particularly on
the ``blue'' CCD, with the lowest S/N near the edges of the orders.
For each spectrum, the continuum was fitted with a 2nd order polynomial, after masking the emission regions.
The fit was then divided into the spectrum, and the remaining continuum subtracted. 
The average processed spectrum for H$\alpha$ is shown in Fig. \ref{ha_avg}, and that for the Bowen blend and \ion{He}{ii}
$\lambda$4686 \AA\ is in Fig. \ref{bowen_heII_avg}. Note that the data in Fig. \ref{bowen_heII_avg} has been
boxcar smoothed, and shifted to the rest frame of the system.
From the averaged spectra, the equivalent width 
for H$\alpha$ is $-3.1$ $\pm$ 0.3 \AA, for \ion{He}{ii} $\lambda$4686 \AA\ is $-2.5$ $\pm$ 0.3 \AA, and for the
Bowen blend is $-2.1$ $\pm$ 0.2 \AA. This is in contrast to the observations of \cite{steeghs2005}, who report that
although their spectroscopy was contaminated by a nearby star, the main feature observed was a broad emission line from
\ion{He}{ii} $\lambda$4686 \AA\, and Balmer emission was at most very weak.

\section{Results}

\subsection{$R$-band Light Curves}

The three light curves in Fig. \ref{lc} 
exhibit a sinusoidal modulation, classically attributed to X-ray irradiation of the inner face
of the secondary, where the modulation is due to the phase dependent visibility of this higher temperature region.
The phasing of the ATT light curve is approximately consistent with this scenario, with the brightest measurements
occurring near phase 0.5. However, the phasing of the other light curves is completely different, suggesting that
the observed modulation does not vary on the orbital period. In particular, the two LBT light curves are separated by
only 11 nights, and in that time the phasing of the light curves, folded on the orbital period, differs
by $\sim$0.2.

\subsection{H$\alpha$ emission}

Fig. \ref{ha_avg} shows the average H$\alpha$ spectrum. This broad emission line exhibits a hint of a double peak
(a classic signature of an accretion disc), although the peaks are not as distinctive as those seen in the spectra of
other LMXBs. The distribution of the flux appears to be symmetrical about a central wavelength, suggesting that, on
average, the emission from the disc is axisymmetrical. To constrain the systemic velocity, $\gamma$, we fitted a
Gaussian to the wings of the averaged spectrum, where the emission is from disc regions closer to the centre, and
presumably less susceptible to corruption from the hotspot and other non-uniformities in the outer disc
\citep{marsh1998}. This gives a value for $\gamma$ of 132 $\pm$ 8 km~s$^{-1}$.

\begin{figure}
  \begin{center}
    \includegraphics[scale=0.39, angle=270]{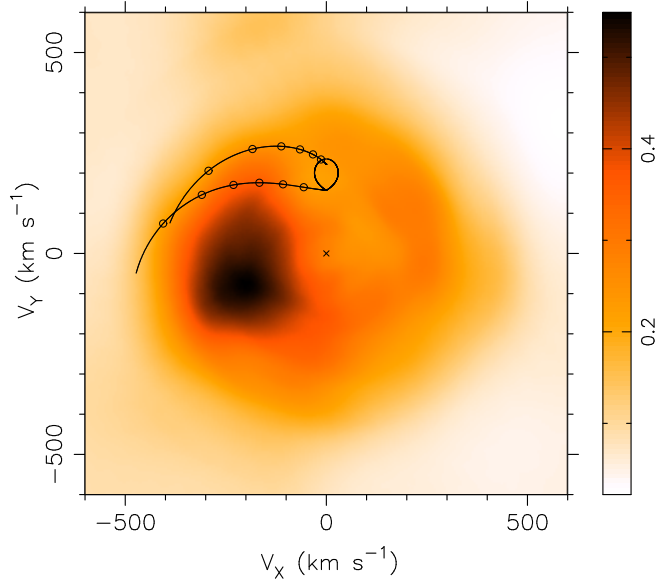}
  \caption{Doppler tomogram for H$\alpha$. The Roche lobe of the secondary is
           plotted. The lower curve represents the gas stream velocity (the
           free-fall ballistic trajectory from the inner Lagrangian point, $L_1$)
           and the upper curve is the Keplerian velocity of the accretion disc along
           the gas stream. Both curves are marked with a circle when the gas stream moves
           0.1R$\mathrm{_{L1}}$ (0.1 of the distance between the primary and $L_1$) closer to
           the primary. The gas stream trajectory and disc velocity along the stream
           were calculated using $K_2 = 200$ km~s$^{-1}$ and $q = 0.035$.}
  \label{ha}
  \end{center}
\end{figure}

We investigated the emission line distribution of the system by examining a Doppler tomogram of the H$\alpha$ line (Fig.
\ref{ha}), constructed using $\gamma = 132$ km~s$^{-1}$. As with the photometry, the spectra were phased using the pulse
timing ephemeris.
For this work, we used the maximum entropy method, implemented in \doppler: see
\cite{marsh1988}, \cite{marsh2001} and \cite{steeghs2004} for further details and applications of Doppler tomography.

Enhanced emission is present in the ($-V\mathrm{_{x}}$,$+V\mathrm{_{y}}$) quadrant, extending into the
(-$V\mathrm{_{x}}$,-$V\mathrm{_{y}}$) quadrant, with a faint ring of
emission. Such a ring of emission is seen in Doppler tomograms of many LMXBs, and is attributed to emission from
the outer disc \citep{marsh1998}.
This emission ring lies between radii of $\sim$200--400 km~s$^{-1}$. Plotting only the axisymmetric component of
this tomogram reveals that the brightest emission is from an annulus at a velocity of $\sim$230 km~s$^{-1}$, but
with significant emission extending to $\sim$150 km~s$^{-1}$. Since the secondary must be moving slower than the
outer disc, we choose 200 km s$^{-1}$ as an upper limit to $K_2$, the velocity of the secondary projected onto the
line of sight.

For illustration, we plot in Fig. \ref{ha} the Roche lobe of the secondary, the gas stream trajectory and accretion disc
velocity curve, computed using $K_2 = 200$ km~s$^{-1}$. Since $K_1$ is already accurately known from the pulse
timing analysis
\cite[$2\pi a_{1}\mathrm{sin}i/P_{\mathrm{orb}} = 6.942 \pm 0.003$ km~s$^{-1}$;][]{kaaret2006},
$q$ ($\equiv K_{1}/K_{2}$)$ = 0.035$. $K_2$ and $q$ are the
only parameters upon which the locations of the Roche lobe and the gas stream trajectory depend. At this value of $K_2$,
the predicted gas stream misses the hotspot location. Lower values of $K_2$ would cause the gas stream
to be closer to the hotspot. 
However, we also note that Doppler tomograms of the \ion{He}{ii} $\lambda$4686 \AA\ emission line in the two NS LMXBs
V801 Ara and V926 Sco also show the peak of emission in the ($-V_{\mathrm{x}}$,$-V_{\mathrm{y}}$) quadrant, close to
the $-V_{\mathrm{x}}$ axis, and for these systems this emission is attributed to an extended disc bulge
\citep{casares2006}. As will be discussed in section 4, the location of the hotspot in
HETE~J1900.1$-$2455 may be explained by a warped, irradiated accretion disc.

\subsubsection{Projected secondary velocity vs. inclination}

\begin{figure}
  \begin{center}
    \includegraphics[scale=0.49, angle=270]{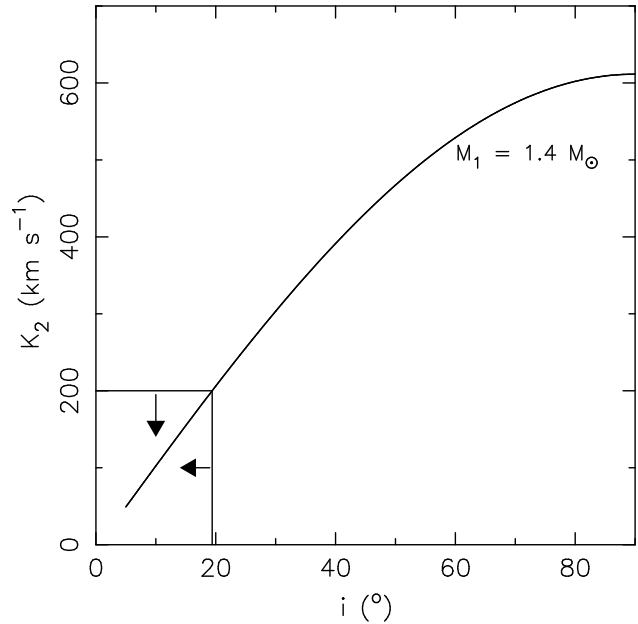}
  \caption{Projected secondary velocity ($K_2$) vs. orbital inclination angle ($i$) for a primary mass of 1.4
           M$_{\sun}$. The upper limit on $K_2$ is set by the velocity of the H$\alpha$ emission ring.}
  \label{kvi}
  \end{center}
\end{figure}

Using the standard mass function equation, and with $K_{2} = K_{1}/q = 2\pi a_{1}\mathrm{sin}i/qP_{\mathrm{orb}}$
it is possible to calculate $q$ and $K_2$ for various orbital inclinations ($i$) and primary masses ($M_1$).
Fig. \ref{kvi} shows a plot of $K_2$ vs. $i$ for a primary mass of 1.4 M$_{\sun}$.
If $K_2$ $<$ 200 km s$^{-1}$, $i$ $\lesssim$ 20\degr, for all $M_1$ $>$ 1.4 M$_{\sun}$.

Hence our upper limit on $K_2$ provides an upper limit on $i$ of $\sim$20$\degr$. This low inclination is
consistent with the fact that the double peak in H$\alpha$ is not as clear as in systems with higher inclinations,
and thus higher projected disc velocities \citep[see for example the LMXB XTE J1118$+$480,
with $i$ $\sim$70\degr;][]{elebert2006}.

\subsection{Bowen blend emission}

\cite{steeghs2002} discovered that emission from the Bowen blend (a blend of
\ion{N}{iii}, \ion{C}{iii} and \ion{O}{ii} lines, centred at $\sim$$\lambda$4640 \AA) traced the motion of
the irradiated secondary star in the LMXB Sco X-1.
Some of the major features seen in the Bowen blend in X-ray binaries are the \ion{N}{iii} lines at
$\lambda$4634.13 \AA, $\lambda$4640.64 \AA\ and
$\lambda\lambda$4641.85/4641.96 \AA, and the \ion{C}{iii} lines at $\lambda$4647.4 \AA, $\lambda$4650.1 \AA\ and
$\lambda$4651 \AA\ \citep{hynes2001,steeghs2002,nelemans2006}.
Our average spectrum of the Bowen blend reveals a broad range of emission between $\lambda$4620 \AA\ and
$\lambda$4650 \AA\ (Fig. \ref{bowen_heII_avg}).
Superimposed on this broad emission, there appear to be several narrower emission features, three of which
correspond to the expected wavelengths of the \ion{N}{iii} $\lambda$4634.13 \AA\ line,
the \ion{N}{iii} blend near $\lambda$4641 \AA, and the $\lambda$4647.4 \AA\ \ion{C}{iii} line.

Because of the low S/N, we performed Doppler tomography of all of the lines in the Bowen blend simultaneously,
with the intention of finding emission from the irradiated face of the secondary. Although this map contains a
single sharp spot of emission, it is in the ($+V_{\mathrm{x}}$,$+V_{\mathrm{y}}$) quadrant, $\sim$45\degr\ away
from the expected location of secondary emission on the $+V_{\mathrm{y}}$ axis. The significance of this
emission is uncertain because of the low S/N of the individual exposures.

\subsection{He {\small{II}} $\lambda$4686 \AA\ emission}

The \ion{He}{ii} $\lambda$4686 \AA\ emission is contaminated in the wings, particularly on the red side. While the
lower S/N in the wings might explain some of this variability, is does not account for all of the systematic artefacts
seen -- in particular, the 2 dips on the red side, near $\lambda$4704 \AA\ and $\lambda$4710 \AA, can be seen in
averages of just a few spectra, and so are not due to
random noise. As well as this, the average line appears asymmetrical, with an excess of flux between
$\lambda\lambda$4695--4705 \AA. The S/N ratio throughout the order is also lower than that for H$\alpha$. Despite this,
a Gaussian fit to the wings of this line (masking the line core, and the excess on the red side) gives a systemic
velocity consistent with that seen in the H$\alpha$ emission.


\section{Discussion}

With its maximum brightness near phase 0.5, we attempted to fit the first of our light curves (Fig. \ref{lc}(a))
using the Eclipsing Light Curve (\elc) code
\citep{orosz2000}, under the assumption that the modulation was due to X-ray heating. This provided
a satisfactory fit, although because of the degeneracy between inclination angle, disc temperature
and disc flaring angle, it was not possible to extract a unique parameter set, and a wide range of
parameter values yielded equally good fits.

However, the phasing of the subsequent light curves (Fig. \ref{lc}(b),(c)) suggests that the observed modulation is not
due to X-ray heating of the secondary: the most likely explanation for this is that the observed modulation is due to
superhumps (see below).
Therefore the apparent alignment of the first light curve is most likely coincidental, and the ELC code (which
does not incorporate superhumps) cannot be used to extract reliable system parameters.

Superhumps are periodic optical modulations, originally seen during
superoutbursts of SU UMa dwarf novae (DNe), but also seen in some LMXBs in outburst
\citep[][and references therein]{haswell2001a}. The phenomenon has also been
seen in persistently bright systems. The superhump period is typically
a few percent longer than $P_{\mathrm{orb}}$. The current understanding is that in systems where the mass ratio is
less than $\sim$0.25, the tidal truncation radius lies outside the 3:1 resonance radius. When the accretion disc
radius extends out to beyond the 3:1 resonance radius, the disc becomes eccentric, and the perturbation from the
secondary causes a prograde precession.
For precessing discs in LMXBs, \cite{haswell2001a} proposed that the disc area changes on the superhump period,
causing the area of the disc intercepting the X-rays to increase.
The increased reprocessing causes the disc to brighten, giving rise to the observed superhumps in LMXBs.
Several systems exhibit superhumps with almost exact sinusoidal modulations like those seen in HETE~J1900.1$-$2455, in
particular another LMXB, XTE J1118$+$480 \citep{patterson2005}.
The peak-to-peak amplitude of the modulation seen in HETE~J1900.1$-$2455
was $\sim$0.1 mag on 2006 September 21 and 2007 June 13, but increased to
$\sim$0.3 mag on 2007 June 24.
The latter amplitude is quite high, but as pointed out by \cite{haswell2001b} superhumps due to irradiated discs are
likely to appear more prominent at {\it low} inclinations: hence the large superhump amplitude supports our estimate
of the inclination in section 3.2.1.
For an X-ray luminosity of $\sim$10$^{36}$ erg s$^{-1}$ the disc is likely to also suffer irradiation-driven warping
\citep[][in particular their model number 1]{foulkes2006} and the disc shape continuously changes under the combination
of prograde apsidal precession and retrograde warped precession. Such models may also explain the factor of $\sim$3
increase in modulation amplitude over a relatively short timescale (Haswell, private communication). 

The location of the hotspot in the H$\alpha$ Doppler tomogram may be interpreted in terms of this
irradiated precessing disc model: as the warped
disc precesses, the gas stream/accretion disc impact point varies. How this occurs can be seen clearly in the
\cite{foulkes2006} simulations\footnote{http://physics.open.ac.uk/FHM\_warped\_disc/q0p075.avi}.
These simulations suggest that multi-epoch Doppler tomography of HETE~J1900.1$-$2455 should reveal a hotspot whose
location varies between the ($-V_{\mathrm{x}}$,$+V_{\mathrm{y}}$) and ($-V_{\mathrm{x}}$,$-V_{\mathrm{y}}$) quadrants.

Previous work has shown that, once the superhump period ($P_{\mathrm{sh}}$) is identified, then the period excess,
$\epsilon$ (= ($P_{\mathrm{sh}} - P_{\mathrm{orb}}$)/$P_{\mathrm{orb}}$), can be directly related to the mass ratio
\citep{patterson2005}. A Fourier transform of our photometry, however, does not allow us to identify a unique superhump
period, because our sampling is too sparse. A more comprehensive photometry campaign is required to measure
$P_{\mathrm{sh}}$, and estimate $q$.

Using the empirical relationship between the absolute visual magnitude of an irradiated accretion disc ($M_{\mathrm{v}}$),
the X-ray luminosity of the central source and orbital period \citep{vanparadijs1994, dejong1996}, we find
$M_{\mathrm{v}} = 3.6 \pm 0.5$ mag. At a distance of 4.3 kpc \citep{suzuki2007}, such a disc should have an apparent
magnitude of 16.7 $\pm$ 0.5 mag. This is consistent with the measured $V$-band magnitude of $\sim$17.1 mag, implying
that the optical light, in the $V$-band, is dominated by an X-ray heated accretion disc. Thus, X-ray heating of the
disc can explain the observed optical brightness of HETE~J1900.1$-$2455.

\cite{kaaret2006} estimated the mass of the secondary to be between 0.016 and 0.07
M$_{\sun}$, for primary masses between 1.4 and 2.2 M$_{\sun}$, but this was based on an assumption of
inclination angle: 90\degr\ for the lower mass limit, and a uniform a priori distribution for the 95 per cent confidence upper
limit. We have seen that for $K_2$ $<$ 200 km~s$^{-1}$, $i\ \lesssim$ 20\degr\ (see Fig. \ref{kvi}).
At this upper limit of $K_2$, $q = 0.035$, and the minimum value of $M_2$ (for $M_1 = 1.4$ M$_{\sun}$) is
therefore 0.05 M$_{\sun}$.

\cite{patterson2003} have computed a mass-radius relation for the secondaries in short period cataclysmic variables
(CVs), finding that in general, the mass-radius curve lies above that for zero-age main sequence stars. An empirical
equation relates $M_2$ to $P_{\mathrm{orb}}$: $M_2 = 0.0764 P_{\mathrm{orb}}^{1.37}\alpha^{-2.05}$~M$_{\sun}$, where
$\alpha$ is the factor relating the radius of the secondaries in these CVs to that of main sequence stars with the
same mass, and $P_{\mathrm{orb}}$ is in hours.
\cite{patterson2003} find that $\alpha$ $\simeq$ 1.18. Assuming that the secondary in HETE~J1900.1$-$2455 is similar
to the secondary in these CV systems, this equation predicts $M_2 = 0.085$ M$_{\sun}$. With the lower limit on $q$
of 0.035, this sets an upper limit to the primary mass of $M_1 < 2.4$ M$_{\sun}$. The secondary may also be a
brown dwarf, in which case $M_2$ is poorly constrained, as is $M_1$.

Finally, we briefly examine the radial velocity of HETE~J1900.1$-$2455 relative to the local Galactic radial
velocity. Based on its position and estimated distance, HETE~J1900.1$-$2455 is $\sim$3.6 kpc from the
Galactic centre, $\sim$1 kpc below the Galactic plane. Using the Galactic rotation curve of \cite{clemens1985}, we
find that the radial velocity of HETE~J1900.1$-$2455 is $\sim$90 km s$^{-1}$ relative to the local Galactic rotation.
Combined with future proper motion studies, this velocity may be used to derive the space velocity of HETE~J1900.1$-$2455
and compute its orbit around the Galactic Centre \citep[e.g.][]{mirabel2003}.


\section{Conclusions}

We have presented a detailed optical study of the AMSP HETE~J1900.1$-$2455. The $R$-band light curves reveal a modulation
whose phase is inconsistent with simple heating of the secondary star/accretion disc: we interpret this as due to
superhumps associated with a precessing accretion disc. Fitting to the wings of the averaged H$\alpha$ emission line
gives the systemic velocity of 132~$\pm$~8~km~s$^{-1}$, and a similar value is found from the \ion{He}{ii} line.
 
Doppler tomography of the H$\alpha$ emission line suggests a upper limit of $\sim$200~km~s$^{-1}$ on the projected
secondary velocity, implying that HETE~J1900.1$-$2455 has a low inclination, $\lesssim$ 20\degr, and a minimum
secondary mass of 0.05~M$_{\sun}$.

Assuming that the secondary in HETE~J1900.1$-$2455 is similar to those seen in short period CVs, we estimate a secondary
mass of 0.085~M$_{\sun}$, which implies an upper limit to the NS mass of 2.4 M$_{\sun}$ (again for
$K_2$ $\lesssim$ 200~km~s$^{-1}$).
If the secondary is a brown dwarf, then this constraint is relaxed.


\section*{Acknowledgments}

We acknowledge the use of the 2.3~m Advanced Technology Telescope of the Australian National University at Siding
Spring Observatory.
The WIYN Observatory is a joint facility of the University of Wisconsin-Madison, Indiana University, Yale University,
and the National Optical Astronomy Observatory.
Some of the data presented herein were acquired using the Large Binocular Telescope (LBT). The LBT is an international
collaboration among institutions in the United States, Italy and Germany. LBT Corporation partners are: The University
of Arizona on behalf of the Arizona university system; Istituto Nazionale di Astrofisica, Italy; LBT
Beteiligungsgesellschaft, Germany, representing the Max-Planck Society, the Astrophysical Institute Potsdam, and
Heidelberg University; The Ohio State University, and The Research Corporation, on behalf of The University of Notre
Dame, University of Minnesota and University of Virginia. We thank LBT director R. Green for allocation of DD time,
and D. Thompson, LBTO for acquiring some of the data.
The spectroscopic data presented herein were obtained at the W.~M. Keck Observatory, which is operated as a scientific
partnership among the California Institute of Technology, the University of California and the National Aeronautics
and Space Administration. The observatory was made possible by the generous financial support of the W.~M. Keck
Foundation.
This research made use of NASA's Astrophysics Data System, and the SIMBAD database, operated at CDS, Strasbourg, France.
X-ray quick-look results provided by the ASM/RXTE team.
We acknowledge the use of \molly\ and \doppler\ software packages developed by T.~R. Marsh, University of Warwick.
We thank J.~X. Prochaska, UCO/Lick Observatory, for reducing our Keck spectroscopy using \redux.
PE and PJC acknowledge support from Science Foundation Ireland.

\appendix

\bsp

\label{lastpage}


\begin{thebibliography}{99}

\bibitem[\protect\citeauthoryear{}{Campana, Cucchiara \& Burrows}{2005}]{campana2005}
   Campana S., Cucchiara A., Burrows D.~N.,
   2005, ATel, 543

\bibitem[\protect\citeauthoryear{}{Casares et\ al.}{2004}]{casares2004}
   Casares J., Steeghs D., Hynes R.~I., Charles P.~A., Cornelisse R., O'Brien K., 
   2004, RevMexAA Conference Series, 20, 21

\bibitem[\protect\citeauthoryear{}{Casares et\ al.}{2006}]{casares2006}
   Casares J., Cornelisse R., Steeghs D., Charles P.~A., Hynes R.~I.,
   O'Brien, K., Strohmayer, T.~E.,
   2006, MNRAS, 373, 1235

\bibitem[\protect\citeauthoryear{}{Clemens}{1985}]{clemens1985}
   Clemens, D.~P,
   1985, ApJ, 295, 422

\bibitem[\protect\citeauthoryear{}{Degenaar et\ al.}{2007a}]{degenaar2007a}
   Degenaar N., Wijnands R., Galloway D.~K., Lewin W.~H.~G., Homan J., Chakrabarty D.,
   Campana S., Miller J.~M., Cackett E.~M.,
   2007a, ATel, 1091

\bibitem[\protect\citeauthoryear{}{Degenaar et\ al.}{2007b}]{degenaar2007b}
   Degenaar N., Campana S., Galloway D.~K., Lewin W.~H.~G., Homan J., Chakrabarty D.,
   Morgan E.~H., Jonker P.~G., Cackett E.~M., Miller J.~M., Wijnands R.,
   2007b, ATel, 1106

\bibitem[\protect\citeauthoryear{}{de Jong, van Paradijs \& Augusteijn}{1996}]{dejong1996}
   de Jong J.~A., van Paradijs J., Augusteijn T.,
   1996, A\&A, 314, 484

\bibitem[\protect\citeauthoryear{}{Dubus, Hameury \& Lasota}{2001}]{dubus2001}
   Dubus G., Hameury J.-M., Lasota J.-P.,
   2001, A\&A, 373, 251

\bibitem[\protect\citeauthoryear{}{Elebert, Callanan \& Torres}{2006}]{elebert2006}
   Elebert P., Callanan P.~J., Torres M.~A.~P.,
   2006, in Meurs E.~J.~A., Fabbiano G., Eds., Proceedings of the 230th Symposium of the International
   Astronomical Union, Cambridge University Press, pp. 57--58

\bibitem[\protect\citeauthoryear{}{Falanga et\ al.}{2005}]{falanga2005}
   Falanga M., Kuiper L., Poutanen J., Bonning E.~W., 
   Di Salvo T., Goldoni P., Goldwurm A., Shaw S.~E., Stella L.,
   2005, A\&A, 444, 15

\bibitem[\protect\citeauthoryear{}{Falanga et\ al.}{2007}]{falanga2007}
   Falanga M., Poutanen J., Bonning E.~W., Kuiper L., Bonnet-Bidaud J.~M.,
   Goldwurm A., Hermsen W., Stella L.,
   2007, A\&A, 464, 1069

\bibitem[\protect\citeauthoryear{}{Foulkes, Haswell \& Murray}{2006}]{foulkes2006}
   Foulkes S.~B., Haswell C.~A., Murray J.~R.,
   2006, MNRAS, 366, 1399

\bibitem[\protect\citeauthoryear{}{Fox}{2005}]{fox2005}
   Fox D.~B.,
   2005, ATel, 526

\bibitem[\protect\citeauthoryear{}{Galloway et\ al.}{2006}]{galloway2006}
   Galloway D.~K., Morgan E.~H., Krauss M.~I., Kaaret P., Chakrabarty D.,
   2006, ApJ, 654, L73

\bibitem[\protect\citeauthoryear{}{Galloway et\ al.}{2007}]{galloway2007}
   Galloway D.~K., Morgan E.~H., Chakrabarty D., Kaaret P.,
   2007, ATel, 1086

\bibitem[\protect\citeauthoryear{}{Garnavich et\ al.}{2007}]{garnavich2007}
   Garnavich P.~M., Callanan P.~J., Elebert P., Reynolds M.~T.,
   2007, ATel, 1110


\bibitem[\protect\citeauthoryear{}{Haswell et\ al.}{2001a}]{haswell2001a}
   Haswell C.~A., King A.~R., Murray J.~R., Charles P.~A.,
   2001a, MNRAS, 321, 475

\bibitem[\protect\citeauthoryear{}{Haswell et\ al.}{2001b}]{haswell2001b}
   Haswell C.~A., Rolfe D.~J., King A.~R., Murray J.~R., Charles P.~A.,
   2001b, ApSSS, 276, 41

\bibitem[\protect\citeauthoryear{}{Hynes et\ al.}{2001}]{hynes2001}
   Hynes R.~I., Charles P.~A., Haswell C.~A., Casares J., Zurita C., Serra-Ricart M.,
   2001, MNRAS, 324, 180

\bibitem[\protect\citeauthoryear{}{Kaaret et\ al.}{2006}]{kaaret2006}
   Kaaret P., Morgan E.~H., Vanderspek R., Tomsick J.~A.,
   2006, ApJ, 639, 963

\bibitem[\protect\citeauthoryear{}{Landolt}{1992}]{landolt1992}
   Landolt A.~U.,
   1992, ApJ, 104, 340

\bibitem[\protect\citeauthoryear{}{Markwardt, Krimm \& Swank}{2007}]{markwardt2007}
   Markwardt C.~B., Krimm H.~A., Swank J.~H.,
   2007, ATel, 1108

\bibitem[\protect\citeauthoryear{}{Marsh and Horne}{1988}]{marsh1988}
   Marsh T.~R., Horne K.,
   1988, MNRAS, 236, 269

\bibitem[\protect\citeauthoryear{}{Marsh}{1998}]{marsh1998}
   Marsh T.~R.,
   1998, in Howell S., Kuulkers E., Woodward C., Eds., ASP Conf. Ser. Vol.
   137, Wild Stars in the Old West, ASP, San Francisco, pp. 236--244

\bibitem[\protect\citeauthoryear{}{Marsh}{2001}]{marsh2001}
   Marsh T.~R.,
   2001, in Boffin H. M. J., Steeghs D., Cuypers J., Eds., Astrotomography:
   Indirect Imaging Methods in Observational Astronomy, LNP Series 573, Springer,
   Berlin, pp. 1--26

\bibitem[\protect\citeauthoryear{}{Mirabel \& Rodrigues}{2003}]{mirabel2003}
   Miirabel I.~F., Rodrigues I.,
   2003, A\&A, 398, L25

\bibitem[\protect\citeauthoryear{}{Morgan, Kaaret \& Vanderspek}{2005}]{morgan2005}
   Morgan E.~H., Kaaret P., Vanderspek R.,
   2005, ATel, 523

\bibitem[\protect\citeauthoryear{}{Nelemans, Jonker \& Steeghs}{2006}]{nelemans2006}
   Nelemans G, Jonker P.~G., Steeghs, D.,
   2006, MNRAS, 370, 255

\bibitem[\protect\citeauthoryear{}{Orosz \& Hauschildt}{2000}]{orosz2000}
   Orosz J.~A., Hauschildt P.~H.,
   2000, A\&A, 364, 265

\bibitem[\protect\citeauthoryear{}{Patterson et\ al.}{2003}]{patterson2003}
   Patterson J., Thorstensen J.~R., Kemp J., Skillman D.~R., Vanmunster T., Harvey D.~A.,
   Fried R.~A., Jensen L., Cook L.~M., Rea R., Monard B., McCormick J., Velthuis F., Walker S.,
   Martin B., Bolt G., Pavlenko E., O'Donoghue D., Gunn J., Nov\'{a}k R., Masi G., Garradd G.,
   Butterworth N., Krajci T., Foote J., Beshore E.,
   2003, PASP, 115, 1308

\bibitem[\protect\citeauthoryear{}{Patterson et\ al.}{2005}]{patterson2005}
   Patterson J., Kemp J., Harvey D.~A., Fried R.~E., Rea R., Monard B., Cook L.~M., Skillman D.~R.,
   Vanmunster T., Bolt G., Armstrong E., McCormick J., Krajci T., Jensen L., Gunn J., Butterworth N.,
   Foote J., Bos M., Masi G., Warhurst P.,
   2005, PASP, 117, 1204

\bibitem[\protect\citeauthoryear{}{Poutanen}{2006}]{poutanen2006}
   Poutanen J.,
   2006, AdSpR, 38, 2697

\bibitem[\protect\citeauthoryear{}{Ragazzoni et\ al.}{2006}]{ragazzoni2006}
   Ragazzoni R., Giallongo E., Pasian F., Baruffolo A., Bertram R., Diolaiti E., Di Paola A., Farinato J.,
   Gentile G., Hill J., Lombini M., Pedichini F., Speziali R., Smareglia R., Vernet E.,
   2006, in Stepp, L.~M., Ed., Society of Photo-Optical Instrumentation Engineers
   (SPIE) Conference, Volume 6267, Ground-based and Airborne Telescopes

\bibitem[\protect\citeauthoryear{}{Savage \& Mathis}{1979}]{savage1979}
   Savage B. D., Mathis J.~S., 
   1979, ARA\&A, 17, 73

\bibitem[\protect\citeauthoryear{}{Steeghs}{2004}]{steeghs2004}
   Steeghs D.,
   2004, AN, 325, 185

\bibitem[\protect\citeauthoryear{}{Steeghs \& Casares}{2002}]{steeghs2002}
   Steeghs D., Casares J.,
   2002, ApJ, 568, 273

\bibitem[\protect\citeauthoryear{}{Steeghs et al.}{2005}]{steeghs2005}
   Steeghs D., Torres M.~A.~P., Garcia M.~R., McClintock J.~E., Miller J.~M.,
   Jonker P.~G., Callanan P.~J., Zhao P., Berlind P., Hutchins R., Watson C.,
   2005, ATel, 543

\bibitem[\protect\citeauthoryear{}{Stetson}{1992}]{stetson1992}
   Stetson P. B.,
   1992, in Worrall D.~M., Biemesderfer C., Barnes J., Eds.,
   Astronomical Data Analysis Software and Systems I, ASP Conf.
   Ser., 25, 297, pp. 297--306

\bibitem[\protect\citeauthoryear{}{Suzuki et\ al.}{2007}]{suzuki2007}
   Suzuki M., Kawai N., Tamagawa T., Yoshida A., Nakagawa Y.~E., Tanaka K.,
   Shirasaki Y., Matsuoka M., Ricker G.~R., Vanderspek R., Butler N., Lamb D.~Q.,
   Graziani C., Pizzichini G., Sato R., Arimoto M., Kotoku J., Maetou M., Yamauchi M.,
   2007, PASJ, 59, 263

\bibitem[\protect\citeauthoryear{}{Torres et\ al.}{2007}]{torres2007}
   Torres M.~A.~P., Rodriguez-Gil P., Steeghs D., Corral-Santana J.~M., Casares J., Jonker P.~G.,
   2007, ATel, 1090

\bibitem[\protect\citeauthoryear{}{Vanderspek et\ al.}{2005}]{vanderspek2005}
   Vanderspek R., Morgan E.~H., Crew G., Graziani C., Suzuki M.,
   2005, ATel, 516

\bibitem[\protect\citeauthoryear{}{van Paradijs \& McClintock}{1994}]{vanparadijs1994}
   van Paradijs J., McClintock J.~E.,
   1994, A\&A, 290, 133

\bibitem[\protect\citeauthoryear{}{Vogt et\ al.}{1994}]{vogt1994}
   Vogt S.~S., Allen S.~L., Bigelow B.~C., Bresee L., 
   Brown B., Cantrall T., Conrad A., Couture M., 
   Delaney C., Epps H.~W., Hilyard D., Hilyard D.~F., 
   Horn E., Jern N., Kanto D., Keane M.~J., 
   Kibrick R.~I., Lewis J.~W., Osborne J., Pardeilhan G.~H., 
   Pfister T., Ricketts T., Robinson L.~B., Stover R.~J., 
   Tucker D., Ward J., Wei M.~Z.,
   1994, SPIE, 2198, 362

\bibitem[\protect\citeauthoryear{}{Wang et\ al.}{2001}]{wang2001}
   Wang Z., Chakrabarty D., Roche P., Charles P.~A., Kuulkers E.,
   Shahbaz T., Simpson C., Forbes D.~A., Helsdon S.~F.,
   2001, ApJ, 563, L61

\bibitem[\protect\citeauthoryear{}{Wijnands}{2006}]{wijnands2006}
   Wijnands R.,
   2006, in Lowry, J.~A., Ed., Trends in Pulsar Research, Nova Science Publishers, New York, pp. 53--78

\bibitem[\protect\citeauthoryear{}{Wijnands \& Van der Klis}{1998}]{wijnands1998}
   Wijnands R., van der Klis, M.,
   1998, Nature, 294, 344

\end{thebibliography}
\end{document}